# Temporal patterns of dispersal-induced synchronization in population dynamics

**Sungwoo Ahn[1], Leonid L Rubchinsky[2,3]**

[1]Department of Mathematics, East Carolina University, Greenville, NC
[2]Department of Mathematical Sciences, Indiana University Purdue University Indianapolis, Indianapolis, IN
[3]Stark Neurosciences Research Institute, Indiana University School of Medicine, Indianapolis, IN

**Highlights**
- Predator-prey oscillating populations with weak dispersal exhibit intermittent synchrony
- Temporal patterns of this synchrony depend on the properties of predator-prey interactions in individual patches and may be independent of synchrony strength
- The study identifies the properties of the predator-prey oscillators responsible for dynamics with numerous short desynchronizations vs. few long desynchronizations

**Abstract**

The mechanisms and properties of synchronization of oscillating ecological populations attract attention because it is a fairly common phenomenon and because spatial synchrony may elevate a risk of extinction and may lead to other environmental impacts. Conditions for stable synchronization in a system of linearly coupled predator-prey oscillators have been considered in the past. However, the spatial dispersion coupling may be relatively weak and may not necessarily lead to a stable, complete synchrony. If the coupling between oscillators is too weak to induce a stable synchrony, oscillators may be engaged into intermittent synchrony, when episodes of synchronized dynamics are interspersed with the episodes of desynchronized dynamics. In the present study we consider the temporal patterning of this kind of intermittent synchronized dynamics in a system of two dispersal-coupled Rosenzweig-MacArthur predator-prey oscillators. We consider the properties of the distributions of durations of desynchronized intervals and their dependence on the model parameters. We show that the temporal patterning of synchronous dynamics (an ecological network phenomenon) may depend on the properties of individual predator-prey patch (individual oscillator) and may vary independently of the strength of dispersal. We also show that if the dynamics of predator is slow relative to the dynamics of the prey (a situation that may promote brief but large outbreaks), dispersal-coupled predator-prey oscillating populations exhibit numerous short desynchronizations (as opposed to few long desynchronizations) and may require weaker dispersal in order to reach strong synchrony.

# 1. INTRODUCTION

Synchronization of dynamics of spatially separated populations appears to be a very general phenomenon (see, e.g., (Liebhold et al., 2004) for a review). The mechanisms and properties of synchronized population dynamics attract attention not only because it is a fairly common phenomenon, but also because spatial synchrony may elevate a risk of extinction (Heino et al., 1997; Earn et al., 2000; Johst and Drechsler, 2003) and may lead to other environmental impacts (e.g., a severe impact of pest outbreaks (Liebhold et al., 2012; Petrovskii et al., 2014)).

One of the major mechanisms of the spatial synchrony is a dispersal between populations (reviewed in Liebhold et al., 2004). Spatial synchrony due to dispersal is prominent in the population with substantial cyclic fluctuations of populations (due to nonlinearity of interactions between species; Vasseur and Fox, 2009). Thus, the dispersal-induced spatial synchrony was considered in mathematical modelling as coupled oscillators. In particular, Goldwyn and Hastings (2008, 2009) provided detailed mathematical analysis of how dispersal can induce synchronization of predator-prey communities, including the consideration of the impact of spatial inhomogeneity. These findings fit very well with the general mathematical view of synchronization of oscillators (e.g., Pikovsky et al., 2001). Indeed, the coupling between predator-prey oscillators due to animal migration would correspond to a linear dissipative coupling, very-well known to have synchronizing effect.

These and many other studies of the synchrony in mathematical models are primarily focused on the stable synchronized state and its associated properties. Generally speaking, stable synchrony requires relatively large coupling between the oscillators. Yet, the dispersal between populations appears to be relatively weak (Kot et al., 1996). Thus, the dispersal may not necessarily lead to a stable synchrony. If the coupling between oscillators is too weak to induce a stable synchrony, oscillators may be engaged into intermittent synchrony, when episodes of synchronized dynamics are interspersed with the episodes of nonsynchronized dynamics (see, e.g., Pikovsky et al., 2001, for a description of transitions to synchronization scenarios).

This leads to the question of what kind of dynamics the weak dispersal can induce in the predator-prey systems. The same moderate level of synchrony may be achieved with two markedly different types of dynamics: many short desynchronized episodes or few long desynchronized episodes (as well as a spectrum of possibilities in between these two extremes). Thus, the objective of this study is to investigate the temporal patterning of the intermittent synchrony in the predator-prey oscillators coupled via weak dispersal.

Similar kind of questions have been studied in other types of biological oscillators (e.g., Park and Rubchinsky, 2012; Ahn and Rubchinsky, 2013, 2017; Ahn et al., 2014a,b). We take the model analysis and data analysis techniques used in those studies and apply them to the dispersal-coupled predator-prey oscillators to investigate how the properties of predator-prey interactions affect the temporal patterning of intermittent synchronization in this ecological system.

## 2. METHODS

### 2.1. Model

To consider interactions between two spatially distinct oscillating predator-prey systems we use the Rosenzweig-MacArthur model in two patches with linear difference coupling. We follow the modelling framework of Goldwyn and Hastings (2008, 2009) as these studies provide a mathematically detailed analysis of how the dispersal in heterogeneous predator-prey systems can affect synchronous dynamics (however, unlike those studies we are considering the weak coupling, resulting in partially synchronized dynamics). This model is relatively simple for mathematical and computational analysis, yet it captures complex synchronization phenomena. We will briefly describe the model here.

The populations for prey and predators are described by the variables $V_i$ and $P_i$ where $i \in \{1,2\}$ is the patch. The dynamics of $V_i$ and $P_i$ is given by:

$$\frac{dV_i}{dt} = r_i V_i \left(1 - \frac{V_i}{K_i}\right) - \frac{c_i a_i P_i V_i}{b_i + V_i} + D_{ji}^V V_j - D_{ij}^V V_i \tag{1}$$

$$\frac{dP_i}{dt} = \frac{a_i P_i V_i}{b_i + V_i} - m_i P_i + D_{ji}^P P_j - D_{ij}^P P_i \tag{2}$$

$$i, j = 1,2; i \neq j$$

The growth of the prey in the absence of predation follows the logistic growth with the intrinsic rate $r_i$ and the carrying capacity $K_i$. Predation has a Holling Type II functional response with predation rate $a_i$ and half saturation coefficient $b_i$. The loss of prey due to predation is also proportional to $c_i$ ($c_i > 1$), the ratio of the loss of prey to the gain in predators. The predator has a linear death rate $m_i$. Migration has a linear per capita rate. There is no immigration or emigration out of the system. Coupling parameters $D_{ij}^V$ and $D_{ji}^V$ represent the prey migration from patch $i$ to $j$ and $j$ to $i$, respectively ($D_{ij}^P$ and $D_{ji}^P$ are analogous for predator migration). We consider small values of coupling so that their individual predator-prey oscillators are weakly coupled (in line with observations of Kot et al., 1996). The heterogeneity is modeled via the differences in parameters $a_i, b_i, c_i, r_i, K_i, m_i$ in two patches.

Following (Goldwyn and Hastings, 2009) this model is rescaled by letting the intrinsic parameters $q_1 = q$ and $q_2 = (1 + \sigma_q)q$ where $q = a, b, c, r, K, m$ and the coupling parameters $D_{12}^V = D^V$ and $D_{21}^V = (1 + \sigma_{DV})D^V$ as well as $D_{12}^P = D^P$ and $D_{21}^P = (1 + \sigma_{DP})D^P$. Further rescaling involves

$$v_i = \frac{V_i}{b}, \; p_i = \left(\frac{ac}{rb}\right) P_i, \; \tau = at, \; \alpha = b/K, \; \eta = m/a, \; \varepsilon = a/r, \; d_{ij}^v = D_{ij}^V/a, \; d_{ij}^p = D_{ij}^P/a. \tag{3}$$

This results in the following system of four ordinary differential equations (Goldwyn and Hastings, 2009):

$$\frac{dv_1}{d\tau} = \frac{1}{\varepsilon}\left(v_1(1 - \alpha v_1) - \frac{p_1 v_1}{1 + v_1}\right) + d^v((1 + \sigma_{dv})v_2 - v_1) \tag{4}$$

$$\frac{dv_2}{d\tau} = \frac{1}{\varepsilon}\left((1 + \sigma_r)v_2(1 - (1 - \sigma_k)\alpha v_2) - (1 + \sigma_a + \sigma_c)\left(1 - \frac{\sigma_b}{1 + v_2}\right)\frac{p_2 v_2}{1 + v_2}\right) \tag{5}$$
$$+ d^v(v_1 - (1 + \sigma_{dv})v_2)$$

$$\frac{dp_1}{d\tau} = \frac{p_1 v_1}{1+v_1} - \eta p_1 + d^p((1+\sigma_{dp})p_2 - p_1) \tag{6}$$

$$\frac{dp_2}{d\tau} = (1+\sigma_a)\left(1 - \frac{\sigma_b}{1+v_2}\right)\frac{p_2 v_2}{1+v_2} - (1+\sigma_m)\eta p_2 + d^p(p_1 - (1+\sigma_{dp})p_2). \tag{7}$$

We consider the oscillatory intrinsic dynamics, which occurs when $\alpha < 1$ and $\eta < \frac{1-\alpha}{1+\alpha}$ (Hastings, 1997). We assume that the coupling strength $d^v = d^p = d$ and $\sigma_a = \sigma_b = \sigma_c = \sigma_r = \sigma_K = \sigma_m = \sigma_{dv} = \sigma_{dv} = \sigma$ where $\sigma > 0$ and $d > 0$. When $\sigma > 0$, two uncoupled patches will have different frequencies. We consider the dynamics of coupled two predator-prey oscillators as we vary oscillators' parameters ε, α, η, and the coupling strength $d$. Numerical simulations of the model were performed in XPP software package (Ermentrout, 2002) using adaptive-step fourth order Runge-Kutta method for 20 units of time with 0.00001 step size.

## 2.2. Analysis of the temporal patterns of synchrony

The objective of this study is to explore the temporal patterning of synchronized dynamics. Roughly speaking, the same moderate synchronization level may be achieved with many short desynchronizations or few long desynchronizations (or different possibilities between these two extremes) and the goal is to discriminate between these possibilities. We have recently developed time-series analysis techniques to characterize temporal patterning of synchronous dynamics (Ahn et al., 2011; Rubchinsky et al., 2014) and applied them to several biological oscillators (mostly neural oscillators, e.g. Park et al., 2010; Ahn and Rubchinsky, 2013, 2017; Ahn et al., 2014a,b; Ratnadurai-Giridharan et al., 2016). The data analysis employed here follows these works very closely and is summarized below.

While there are multiple ways to analyze synchronization phenomena, we will use phase-based analysis. Phase synchronization is a common phenomenon in weakly coupled oscillators (Pikovsky et al., 2001). The phase synchronization has been considered in ecological dynamics with model and real data (e.g., Blasius et al., 1999) as well as in the studies of neural dynamics (see references above). For an oscillatory activity, a phase of the $i$th prey-predator system is reconstructed here by computing

$$\varphi_i(t) = \arctan\left(\frac{v_i(t) - \hat{v}_i}{p_i(t) - \hat{p}_i}\right) \tag{8}$$

with atan2-type function to reconstruct angular coordinate, $(\hat{v}_i, \hat{p}_i)$ is a middle point of oscillations in the $(v_i, p_i)$-plane. Then we consider an average synchronization strength index to measure the strength of the phase locking between two signals (e.g., Pikovsky et al., 2001; Hurtado et al., 2004):

$$\gamma = \left\|\frac{1}{N}\sum_{j=1}^{N} e^{i\Delta\varphi(t_j)}\right\|^2 \tag{9}$$

where $\Delta\varphi(t_j) = \varphi_1(t_j) - \varphi_2(t_j)$ is the phase difference, the $t_j$ are the sampling points, $N$ is the number of data points to be considered, and $\|.\|$ is the absolute value of a complex number. This phase synchronization index γ varies from 0 (lack of synchrony) to 1 (perfect synchrony). It provides average value of phase-locking. There may be cycles of oscillations, when phase difference is close to the average value of the phase difference (phase-locked, synchronized state) and when it is not close to it (desynchronized state).

To study the fine temporal structure of the dynamics of synchronization, we construct a sequence of phase lags between signals. Whenever $\varphi_1$ crossed the zero from negative to positive values, we recorded the value of $\varphi_2$, generating a set of consecutive phase values $\{\phi_i\}, i = 1, ..., M$. If the value of $\phi_i$ differs from the average value of $\phi_i$ by less than $\pi/2$ then the oscillations are considered to be in a synchronized state, otherwise they are in the desynchronized state. The choice of $\pi/2$ value for the threshold follows earlier applications of these method (see references above). We used the Kolmogorov-Smirnov test to detect non-uniform distribution of $\{\phi_i\}_{i=1}^{M}$ with the significance level of 0.05 to include it in the further analyses (the results were not qualitatively affected by this level). The duration of desynchronizations is defined as the number of cycles of oscillations that the system spends in the desynchronized state. Note that synchronized state here is the one with near constant (but not necessarily zero) phase lag (which is in line with observations of non-zero lag population synchrony, e.g., Martin et al., 2017).

We characterize the temporal patterning of intermittent synchronization by considering the distribution of desynchronization durations (measured in the cycles of oscillations, thus duration is a discrete variable, as described above). In particular, we consider the mode of this distribution. For example, mode=1 indicates that most common desynchronization duration is very short. We also consider $p_{mode}$ –the probability to observe the duration, which corresponds to the mode (i.e. the chance to observe the most common duration). There is a reason for this approach: if the mode of the desynchronization duration is small, but long desynchronizations are still fairly frequent, then the dynamics is not necessarily dominated by short desynchronizations overall. However, if $p_{mode}$ is relatively large, this guarantees that all other desynchronization durations are rare.

## 3. RESULTS

### 3.1. The impact of ε (the ratio of the predation rate to the intrinsic prey growth rate) on the temporal patterning of synchronization

We consider here how ε affects the durations of desynchronization events. Parameter ε is a ratio of predation rate to the intrinsic rate of the prey growth. In the limit as ε goes to zero, the system becomes a relaxation oscillator. As the value of ε increase, the fine temporal structure of synchronization changes when ε is about 0.1 as evident by the changes of the mode of the distribution of desynchronization durations (Fig. 1A). Smaller values of ε promote short desynchronization episodes that last for only one cycle of oscillations. The increase in ε leads to the increase of the mode of the distribution of desynchronization durations.

There is also an effect on the synchrony strength γ (Fig. 1B). Increasing ε leads to a modest decrease in the synchrony strength γ. The frequency of oscillations is also affected by ε although quite weakly (Fig. 1C). This means that as ε decreases, the mean desynchronization duration is short not only if measured in relative units (cycles of oscillations), but also in the absolute units of time.

Note that the probability of the dominant duration of desynchronization events $p_{mode}$ (the insert in Fig. 1A) is always higher than 0.5. Thus, more than the half of all desynchronizations are captured by the mode (1 or 2 cycles here) and mode of the distribution is really representative of the dynamics in this situation.

## 3.2 The impact of η (the ratio of the predator death rate to predation rate) on the temporal patterning of synchronization

We consider here how $\eta$ affects the synchrony strength and the durations of desynchronization events. Smaller predator death rate (and thus smaller η) increases the amount of time necessary for the predator population to become sufficiently small so as to allow for a prey outbreak, increasing the amount of time in the cycle with low prey population. As the value of η increase, the fine temporal structure of synchronization changes as evident by the changes of the mode of the distribution of desynchronization durations (Fig. 2A). Smaller values of η promote short desynchronization episodes lasting for two cycles of oscillations. On the contrary, the increase in η leads to the increase of the mode of the distribution of desynchronization durations to three and four cycles (that is by a factor of two).

As η changes, there is also an effect on the synchrony strength γ (Fig. 2B). Shorter desynchronizations correspond to the higher synchrony level. Fig. 2C shows that the mean frequency is also almost constant (minor increase). Thus, like for the variation of ε case considered above, the desynchronizations are short here not only if measured in the number of cycles, but also measured in absolute time units. And again, the probability of the dominant duration of desynchronization events $p_{mode}$ (the insert in Fig. 2A) is mostly close to 1 and always higher than 0.5, indicating that the mode captures the majority of desynchronizations.

## 3.3 The impact of α (the ratio of the predation functional response half-saturation to the prey carrying capacity) on the temporal patterning of synchronization

We consider now how α affects the synchrony strength and the durations of desynchronization events. Decreasing α increases the carrying capacity, thereby increasing both the magnitude of the prey outbreak and the time between outbreaks. Smaller values of α promote short desynchronization episodes lasting for two cycles of oscillations. As α increases, the most frequent desynchronization episodes are getting longer (Fig. 3A). As α changes, there is also an effect on the synchrony strength γ in the range of smaller α, which virtually disappears for larger α ; frequency does not depend on α in the considered range of the variation of this parameter (Fig. 3B,C). Similar to what was observed above, the probability of the dominant duration of desynchronization events $p_{mode}$ (the insert in Fig. 3A) is mostly close to 1 and always higher than 0.5, indicating that the mode captures majority of desynchronizations.

## 3.4 Changing desynchronization durations independently of frequency and synchrony strength

Earlier studies showed that one can change parameters of oscillators in such a way that the distribution of desynchronization duration is changed independently of the average synchrony strength (e.g., Ahn et al., 2011; Rubchinsky et al., 2014; Ahn and Rubchinsky, 2017). In the coupled ecological oscillators considered here, changes in the temporal patterning of synchronization can be independent of the synchronization strength too, as evidenced by the results presented above (although they may co-vary together as well). This evidence is somewhat limited, because in the ranges of parameters studied, eventually changes in the synchrony strength are followed by the changes in the synchrony patterns. Nevertheless, when α is in the range of [0.49, 0.54] (Fig. 3), both the synchrony strength $\gamma$ and the mean frequency of oscillators do not vary much (Fig. 3B, C) while the mode of desynchronization durations changes substantially (from cycle 2 to cycle 4, Fig. 3A). This kind of situation is present to a lesser degree in Figs. 1 and 2. The point, however, is that the same level of synchrony strength may be supported either with relatively large number of fairly short desynchronizations or a smaller number of long desynchronizations regardless of whether the durations of desynchronizations are measured in cycles of oscillations or in absolute time units.

### 3.5. Dispersal-induced synchronization threshold for different temporal patterns of intermittent synchronization

We measure the threshold value of the dispersal rate $d$ to reach strongly synchronized dynamics (dynamics without any desynchronization events) between the dynamics of two patches in the parameter regimes as in Fig. 1 (change ε), Fig. 2 (change η), and Fig. 3 (change α). The parameters for each desynchronization duration mode (measured in cycles of oscillations) at each setting were chosen as the smallest parameter to achieve the given mode of the desynchronization durations. While one cannot directly vary the duration of desynchronizations, we vary system parameters to change the mode of the desynchronization durations, which is reflected in the horizontal axes in the Fig. 4. For example, for the Fig. 4A, the system achieves Cycle 1 when $0.08 \leq \varepsilon \leq 0.09$, Cycle 2 when $0.1 \leq \varepsilon \leq 0.15$. Then in Fig. 4A, for Cycle 1 the value of $\varepsilon = 0.08$, for Cycle 2 the value of $\varepsilon = 0.10$. In Fig. 4B, for Cycle 2 the value of $\eta = 0.27$, for Cycle 3 the value of $\eta = 0.29$, for Cycle 4 the value of $\eta = 0.30$. In Fig. 4C, for Cycle 2 the value of $\alpha = 0.47$, for Cycle 3 the value of $\alpha = 0.50$, for Cycle 4 the value of $\alpha = 0.54$. The results presented in Fig. 4 indicate that the system with short desynchronization dynamics needs weaker dispersal strength to be synchronized than the system with longer desynchronization dynamics (even if the initial synchrony strength is nearly the same).

### 4. DISCUSSION

We considered two predator-prey populations coupled via dispersion modeled as two Rosenzweig-MacArthur oscillators with linear difference coupling. Models of individual predator-prey patches are rescaled to get dimensionless variables and have slightly different parameter values to represent spatial heterogeneity following the framework introduced in the studies of Goldwyn and Hastings (2008, 2009). However, unlike those studies (and in line with the observations of dispersion being weak, Kot et al., 1996), we consider the case where the coupling strength is weak enough (relative to the difference of frequencies in isolated oscillators), so that the coupling-induced synchrony is only partial. The dynamics of these systems exhibits intermittent synchrony just due to the moderate values of coupling (without any indigenous noise or environmental stochasticity). This system exhibits intervals of time when the dynamics is synchronized, and intervals of time when the dynamics is desynchronized.

We found that the temporal structure of this dynamics exhibit dependence on several parameters of predator-prey model oscillators. In particular, the larger values of the ratio of predation rate to intrinsic growth rate for prey, the ratio of predator half saturation coefficient to the carrying capacity of prey, and the ratio of predator death rate to the predation rate lead to the dynamics with longer desynchronization episodes (there are ratios of parameters here, because we considered nondimensionalized systems). While changes in model parameters may affect both the average synchrony strength and temporal patterning of synchronized dynamics, the average synchrony strength and its temporal patterning can be independent. The present study shows how the durations of desynchronizations may be altered while the average synchrony strength stays the same. Thus, the same synchrony strength between migration-coupled predator-prey populations may be achieved via many short desynchronized episodes or few long desynchronized episodes.

Our numerical analysis shows that the temporal patterns of synchrony (dynamics of ecological network) may be altered just by the alteration of the properties of predator-prey interactions (properties of individual oscillators) and this may be altered without changes in the dispersal strength. On the other hand, dispersal strength is naturally affecting the overall synchrony strength as larger coupling strength in general leads to complete synchrony between oscillators. What we see here is that the interplay of the individual oscillator properties and dispersal leads to a potentially important observation: dispersal-coupled predator-prey oscillating populations exhibiting fewer longer desynchronizations (as opposed to numerous short desynchronizations) may require stronger dispersal in order to reach strong synchrony. It is known that the weak dispersal may have complex effects on the stability and survival of synchronized predator-prey populations (see, e.g., Abbot, 2011). We show here a case where the same dispersal leads to different impacts on the synchronized dynamics depending on the characteristics of predator-prey interactions.

The changes in the parameter values (towards the smaller values of $\varepsilon$, $\alpha$, and $\eta$) that promote shorter desynchronizations (even if the average synchrony stays the same) essentially make the dynamics of predator slow relatively to the dynamics of the prey, separating the time-scales of predator and prey dynamics. These small parameter values lead to the predator dynamics being initially slow to follow the increase in prey, eventually leading to late but very sharp rise (outbreak) in predator numbers. The predator-prey systems with substantially different timescales and resulting dynamics of outbreaks has long been considered in mathematical ecology (e.g., Ludwig et al., 1978; Rinaldi and Scheffer, 2000). It is worth mentioning that dispersal-synchronized resources enrichment-induced outbreak-like population cycles were found to exhibit low persistence under some conditions in a laboratory experiments (Laan and Fox, 2019), which suggest that mechanisms of outbreaks may exhibit complex interplay with the synchronized dynamics and persistence. Perhaps the outbreak-type oscillatory cycles effectively modulate the coupling making it very strong at the top of the population peak while keeping it low at other times (the coupling strength is constant in this consideration, it is the magnitude of the coupling term that effectively increases, because it is proportional to population difference between patches).

Interestingly, another example of biological oscillator with separation of timescales is a spiking neuron. In particular, many neurons have sodium-potassium spiking mechanism with fast $Na^+$ and slow "delayed rectifier" $K^+$ currents and associated models of Hodgkin-Huxley type have substantially different time scales (see, e.g., Izhikevich, 2007; Ermentrout and Terman, 2010). The separation of the time scales in neuronal models also led to the prevalence of short desynchronizations, if the neurons were connected via weak synapses and exhibited partially synchronized dynamics (Ahn and Rubchinsky, 2017). This probably points to a very general mathematical basis of this phenomenon, which calls for its mathematical exploration.

It would be interesting and important to consider the temporal patterns of synchronization and desynchronization in the more realistic ecological context. Various sources of stochasticity (whether indigenous or environmental) as well as seasonal variability may affect population synchrony (e.g., Bressloff and Lai, 2013) and thus may potentially affect the temporal patterns of synchronized dynamics. In particular, it may be important to consider temporal patterns of ecological synchrony in the context of Moran effect (e.g., Goldwyn and Hastings, 2011). Moreover, various spatial effects may affect synchronous dynamics (Walters et al., 2017, Hopson and Fox, 2019) and may be relevant here as well. Overall, it is important not to overinterpret the results of this study. The parameters of the model are not traced to a specific ecological system. The actual mechanisms of the ecological oscillations and synchrony may involve much more that the ones represented by the predator-prey interactions of Rosenzweig-MacArthur model considered here (see e.g., Barraquand et al., 2017). Thus, this study presents only a potential possibility for the temporal patterns of intermittent synchronous dynamics, its dependence on parameters of predator-prey interaction, and its potential impact on long-term ecological dynamics.

Finally, we would like to put our observations in the context of the studies of transient dynamics in ecology, which appears to be very important and quite common (Hastings et al., 2018) including population synchronization phenomena (Klapwijk et al., 2018). From a mathematical perspective, the idea that the ecologically relevant dynamics is not necessarily the dynamics near and on the attracting synchronization manifold in the phase space is somewhat similar to the idea of importance of transients in ecology. While the intermittent synchronization considered here is not a transient phenomenon in a strict sense, in both situations the properties of dynamics of interest depend not only on the properties of attractors in the phase space, but also on how the system approaches to and leaves them.

## Acknowledgements

L.L.R. would like to acknowledge the support by the National Science Foundation (grant DMS 1813819) and the very useful discussions at the Transients in Biological Systems workshop at the National Institute of Mathematical and Biological Synthesis.

## Conflicts of interest

The authors declare no conflicts of interest.

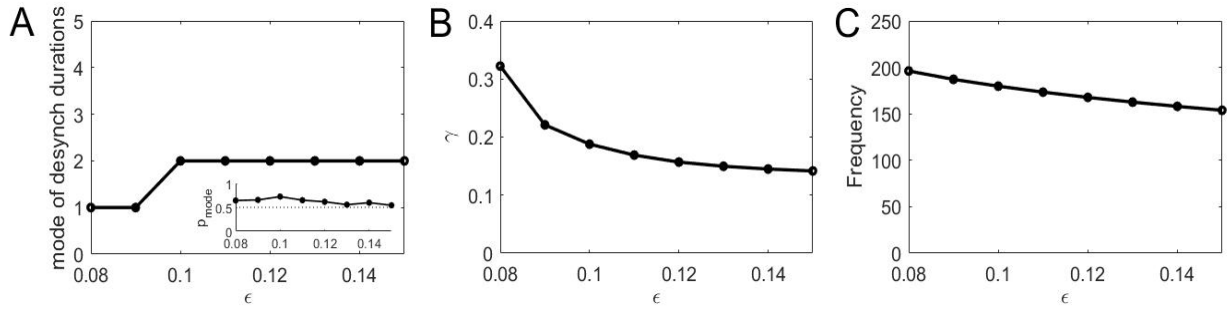

Figure 1. The effect of $\varepsilon$ (the ratio of the predation rate to the intrinsic prey growth rate). (A) Mode value of the durations of desynchronization events (solid line with black dots) and the corresponding probability to observe the mode value $p_{mode}$ (dotted line). The high (close to one) value of probability to oberve a mode indicates that the desynchronizations of corresponding duration are strongly prevalent. (B) Synchronization strength index $\gamma$. (C) The mean frequency of oscillations in coupled predator-prey oscillators. The other parameters are $\alpha = 0.34$, $\eta = 0.32$, $d = 0.03$, ratio $\sigma = 0.19$.

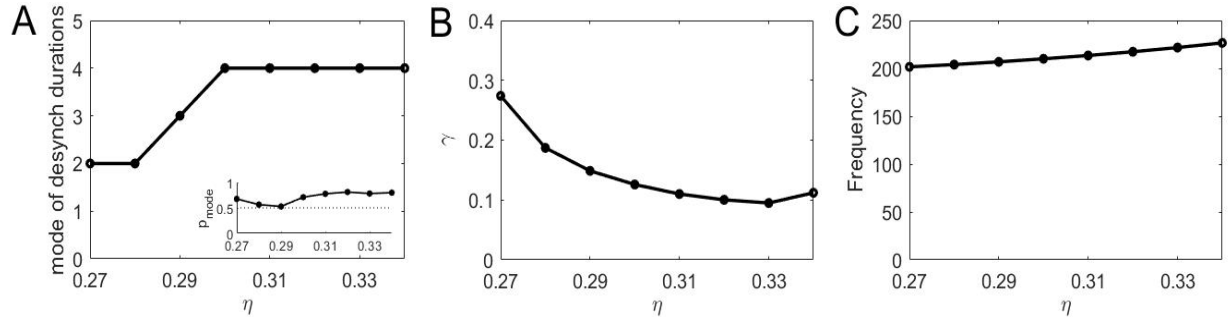

Figure 2. The effect of $\eta$ (the ratio of the predator death rate to predation rate). (A) Mode value of the durations of desynchronization events (black line with black dots) and the corresponding probability to observe the mode value $p_{mode}$ (dotted line). (B) Synchronization strength index $\gamma$. (C) The mean frequency of oscillations in coupled predator-prey oscillators. The other parameters are $\varepsilon = 0.09$, $\alpha = 0.47, d = 0.05$, ratio $\sigma = 0.11$.

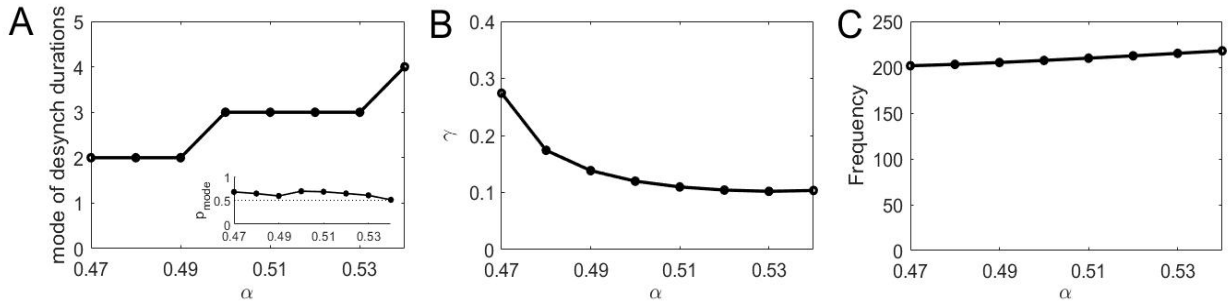

Figure 3. The effect of $\alpha$ (the ratio of the predation functional response half-saturation to the prey carrying capacity). (A) Mode value of the durations of desynchronization events (black line with black dots) and the corresponding probability to observe the mode value $p_{mode}$ (dotted line). (B) Synchronization strength index $\gamma$. (C) The mean frequency of oscillations in coupled predator-prey oscillators. The other parameters are $\varepsilon = 0.09$, $\eta = 0.27$, $d = 0.05$, ratio $\sigma = 0.11$.

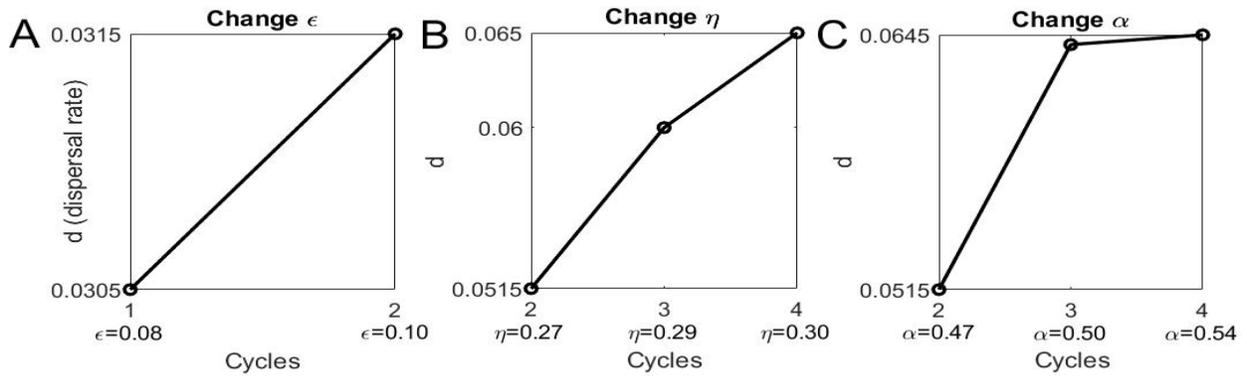

Figure 4. Threshold value of the dispersal rate $d$ to reach dynamics without any desynchronization events for different lengths of cycles by changing (A) $\varepsilon$ (Fig. 1), (B) $\eta$ (Fig. 2), (C) $\alpha$ (Fig. 3). (A) For Cycle 1 the value of $\varepsilon = 0.08$, for Cycle 2 the value of $\varepsilon = 0.10$. (B) For Cycle 2 the value of $\eta = 0.27$, for Cycle 3 the value of $\eta = 0.29$, for Cycle 4 the value of $\eta = 0.30$. (C) For Cycle 2 the value of $\alpha = 0.47$, for Cycle 3 the value of $\alpha = 0.50$, for Cycle 4 the value of $\alpha = 0.54$.